# Ramsey-type phase control of free electron beams

Katharina E. Echternkamp[1], Armin Feist[1], Sascha Schäfer[1,+], and Claus Ropers[1,2,*]

[1] 4th Physical Institute – Solids and Nanostructures, University of Göttingen, Germany
[2] International Center for Advanced Studies of Energy Conversion (ICASEC), University of Göttingen, Germany

**Interference between multiple distinct paths is a defining property of quantum physics,[1] where "paths" may involve actual physical trajectories, as in interferometry,[2] or transitions between different internal (e.g. spin) states,[3] or both.[4] A hallmark of quantum coherent evolution is the possibility to interact with a system multiple times in a phase-preserving manner. This principle underpins powerful multi-dimensional optical[5] and nuclear magnetic resonance[3] spectroscopies and related techniques, including Ramsey's method of separated oscillatory fields[6] used in atomic clocks. Previously established for atomic, molecular and quantum dot systems,[7] recent developments in the optical quantum state preparation of free electron beams[8] suggest a transfer of such concepts to the realm of ultrafast electron imaging and spectroscopy.**

**Here, we demonstrate the sequential coherent interaction of free electron states with two spatially separated, phase-controlled optical near-fields. Ultrashort electron pulses are acted upon in a tailored nanostructure featuring two near-field regions with anisotropic polarization response. The amplitude and relative phase of these two near-fields are independently controlled by the incident polarization state, allowing for constructive and destructive quantum interference of the subsequent interactions. Future implementations of such electron-light interferometers may yield unprecedented access to optically phase-resolved electronic dynamics and dephasing mechanisms with attosecond precision.**

A central objective of attosecond science is the optical control over electron motion in and near atoms, molecules and solids, leading to the generation of attosecond light pulses or the study of static and dynamic properties of bound electronic wavefunctions.[9-13] One of the most elementary forms of optical control is the dressing of free electron states in a periodic field,[14,15] which is observed, for example, in two-color ionization,[16,17] free-free transitions near atoms,[14,18] and in photoemission from surfaces.[19-21] Similarly, beams of free electrons can be manipulated by the interaction with standing waves[22,23] or optical near-fields.[24-27,8] In this process, field localization at nanostructures facilitates the exchange of energy and momentum between free electrons and light. In the past few years, inelastic electron-light scattering[25,26,28] found application in so-called "photon-induced near-field electron microscopy" or PINEM,[24,29,30,8] the characterization of ultrashort electron pulses,[26,27,30] or in work towards optically-driven electron accelerators.[32,33] Very recently, the quantum coherence of such interactions was demonstrated by observing multilevel Rabi-oscillations in the electron populations of the comb of photon sidebands.[8,25] Access to these quantum features, gained by nanoscopic electron sources of high spatial coherence,[34,35] opens up a wide range of possibilities in coherent manipulations, control schemes and interferometry with free electron states.

*Email: cropers@gwdg.de, +Email: schaefer@ph4.physik.uni-goettingen.de

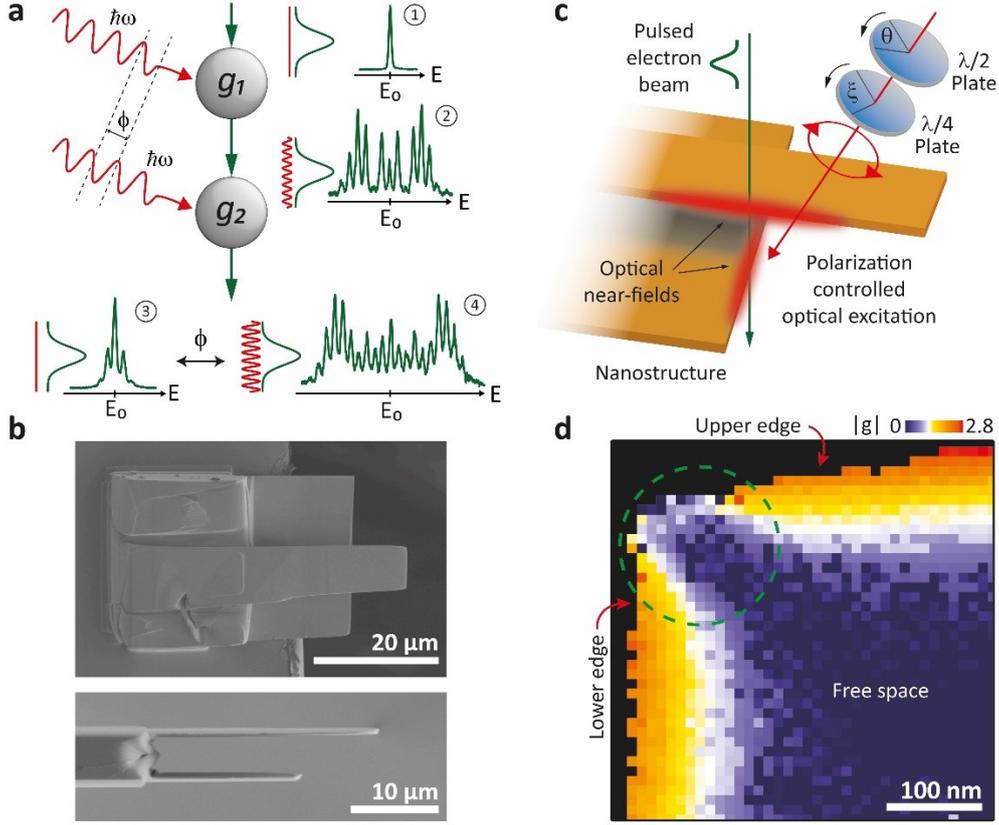

Figure 1: Experimental concept and setup. **a** Working principle of the Ramsey-type free electron interferometer: An electron pulse (green) is acted upon at two spatially separated nodes $g_1$ and $g_2$. A sinusoidal phase modulation is imprinted onto the electron wave function during the first interaction, leading to the generation of spectral sidebands. The relative phase of the interactions governs the phase modulation of the final state. ① to ④: Experimental electron energy spectra. ① incident spectrum. ② Spectrum for a single interaction. ③,④ Spectra recorded for destructive and constructive double interactions, respectively. **b** Scanning electron micrographs of the nanostructure featuring two interaction zones (top and side view). Distance between gold paddles: 5 µm. **c** Sketch of the experimental scenario displaying polarization-controlled excitation of the nanostructure. **d** Raster-scanned image of the local coupling strength $|g_{tot}|$ (see text) for excitation conditions near complete recompression in the corner region (green dashed circle, waveplate angles θ = -38°, ξ = 26°, cf. Fig. 3**b**).

Here, we present a first implementation of quantum coherent sequential interactions with free electron pulses. In particular, we employ a nanostructure that facilitates phase-controlled double interactions, leading to a selectable enhancement or cancellation of the quantum phase modulation in the final electron wavefunction. Figure 1**a** illustrates the basic principle of our approach: Traversal of the first near-field induces photon sidebands (labeled ② in Fig.1**a**) to the initially narrow electron kinetic energy spectrum (①), which correspond to a sinusoidal phase modulation of the free electron wavefunction. Following free propagation, the electrons coherently interact with a second near-field and, in analogy to Ramsey's method,[6] the final electronic state sensitively depends on the relative phase between the two acting fields. In particular, a further broadening (④) or a recompression (③) of the momentum distribution can be achieved.

For a single interaction of a free, quasi-monoenergetic electron state with an optical near-field, the resulting final state is composed of a superposition of momentum sidebands associated with energy changes by ±N photon energies,[25,26] populated with amplitudes $A_N$ according to

$$A_N = \left(\frac{g}{|g|}\right)^N J_N(2|g|), \qquad (1)$$

where $J_N$ are the $N^{th}$-order Bessel functions. The dimensionless coupling parameter $g$ describes the efficiency of momentum exchange with the electron and scales linearly with the incident optical field amplitude (see Supplementary Material S1). In the spatial representation of the free electron state, this Bessel-type distribution of sideband amplitudes is manifest in a sinusoidal modulation of the phase of the wavefunction in the form[26]

$$\psi_{\text{fin}} = \exp\left(2i|g|\sin\left(\frac{\omega z}{v} + \arg(g)\right)\right)\psi_{\text{in}}, \qquad (2)$$

where $\psi_{\text{in}}$ and $\psi_{\text{fin}}$ are the initial and final state wavefunctions, respectively, $\omega$ the optical frequency, $v$ the electron velocity, and $z$ the spatial coordinate along the electron trajectory.

In the present experiment, schematically depicted in Figs. **1a,c**, we demonstrate that two spatially separated optical near-fields may cause an overall interaction of strength $g_{tot}$, which is describable as the coherent sum of the individual, generally complex-valued interactions $g_1$ and $g_2$,

$$g_{tot} = g_1 + e^{i\varphi_0} g_2, \qquad (3)$$

where $\varphi_0$ is a constant phase offset that depends on the spatial separation of the interaction regions (see Supplementary Material S1, S3). In terms of the spatial wavefunction, this then corresponds to an overall enhancement or cancellation of the subsequent interaction-induced phase modulations (eq. 2).

The desired control over $g_{tot}$ requires the ability to separately address the two near-fields in a phase-locked manner. We achieve this by tailoring the nanostructure geometry, employing the strong polarization anisotropy of a pair of perpendicular plates (Fig. **1b**). This approach allows us to control the near-field strengths and their relative phase by selecting the polarization state of the overall excitation. In the following, we describe the experimental implementation of this principle.

A narrow beam of ultrashort electron pulses passes the optically excited nanostructure in close vicinity (Fig. **1c**). The final electronic state resulting from inelastic electron-light scattering is analyzed by electron spectroscopy upon a systematic variation of the incident light polarization. The polarization state is described by the Jones vector ***J***, which we set in the standard fashion[36] by the combination of a half- and quarter wave plate at rotation angles $\theta$ and $\xi$, respectively. The Jones vector for sample excitation is then given by the product of the initial (in our case diagonal) polarization state and wave plate Jones matrices *M*, scaled by the field strength $F$=0.08 V/nm: $\boldsymbol{J} = F \cdot M_{\lambda/4}(\xi) \cdot M_{\lambda/2}(\theta) \cdot \frac{1}{\sqrt{2}} \begin{pmatrix} 1 \\ 1 \end{pmatrix}$.

In a first set of measurements, the near-field responses of the two nanoscopic plates to the incident polarization state are independently characterized. To this end, the electron beam is placed close to each of the edges, and distant from the corner (red, blue circles in Fig. 2**f**), such that the electrons only traverse one of the two near-field regions in each case. Figures 2**a,b** display electron spectra for a continuous variation of polarization states

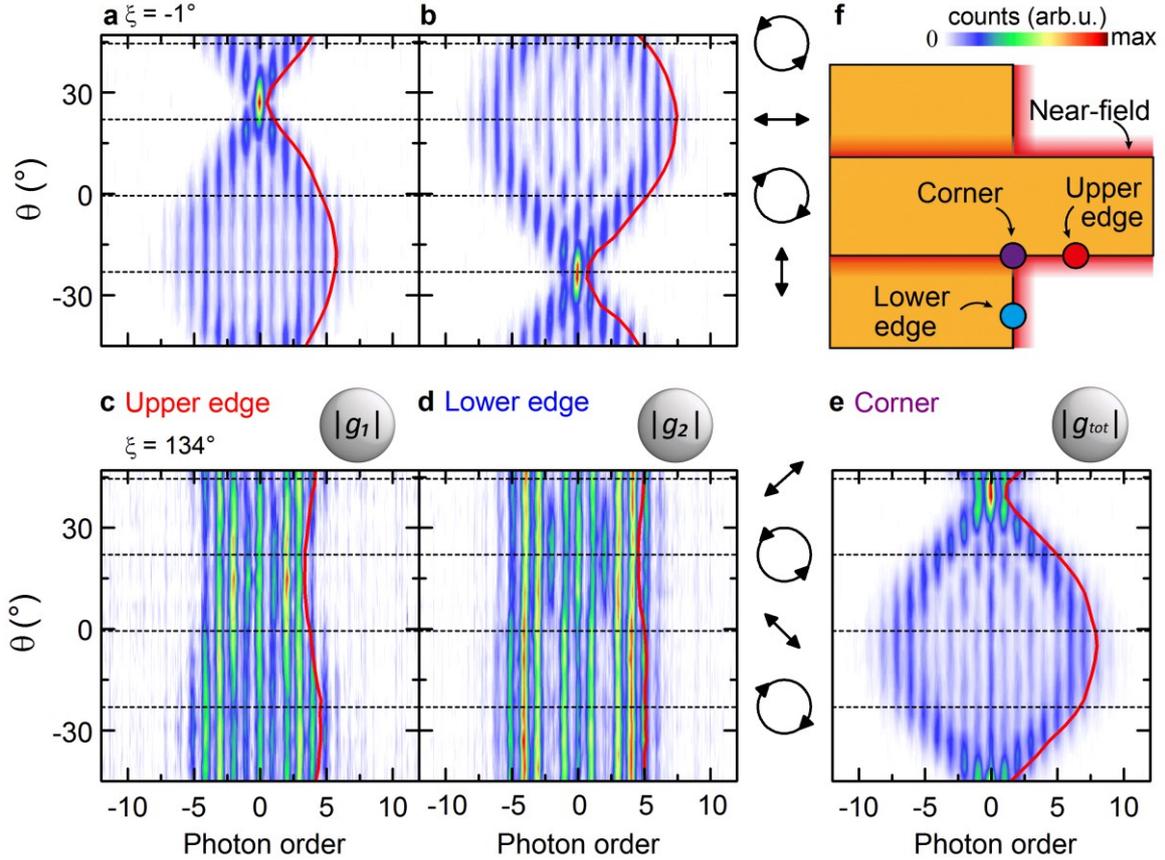

Figure 2: Phase-control of dual interaction. **a-e** Electron spectra recorded at different positions on the sample for varying incident polarization states (arrow icons). Red lines: Coupling constant $2|g_{1,2}|$ extracted from the spectra (see text). **f** Illustration of the three different measurement positions leading to single (red, blue circles) and double interaction (purple). **a,b** The upper (lower) edge yields maximum interaction strength for vertical (horizontal) incident polarization. **c,d** For polarization states varying from diagonal to circular, nearly constant interaction strengths at the individual edges are observed. **e** At the corner, the associated change of relative phase of the interactions $|g_1|$ and $|g_2|$ leads to a strong modulation in $|g_{tot}|$, demonstrating the coherent actions of $g_1$ and $g_2$.

(achieved by wave plate rotation), including polarizations parallel and perpendicular to the plates. The widths of these spectra directly reflect the respective coupling constants $g_{1,2}$, as the highest populated sideband is given by $2|g|$.[8] It is evident that both edges exhibit strong near-fields only for excitation conditions with polarization perpendicular to the respective edge orientation. This behavior can be regarded as a linear analyzer response, in which each edge projects the incident polarization state onto a quasi-polarizability $\alpha_{1,2}$, yielding scalar coupling constants $g_{1,2}=\boldsymbol{\alpha}_{1,2}\cdot\boldsymbol{J}$. By the design of the structure, the vectors $\boldsymbol{\alpha}_{1,2}$ are linearly independent, in fact nearly orthogonal, which allows for separate amplitude and phase control.

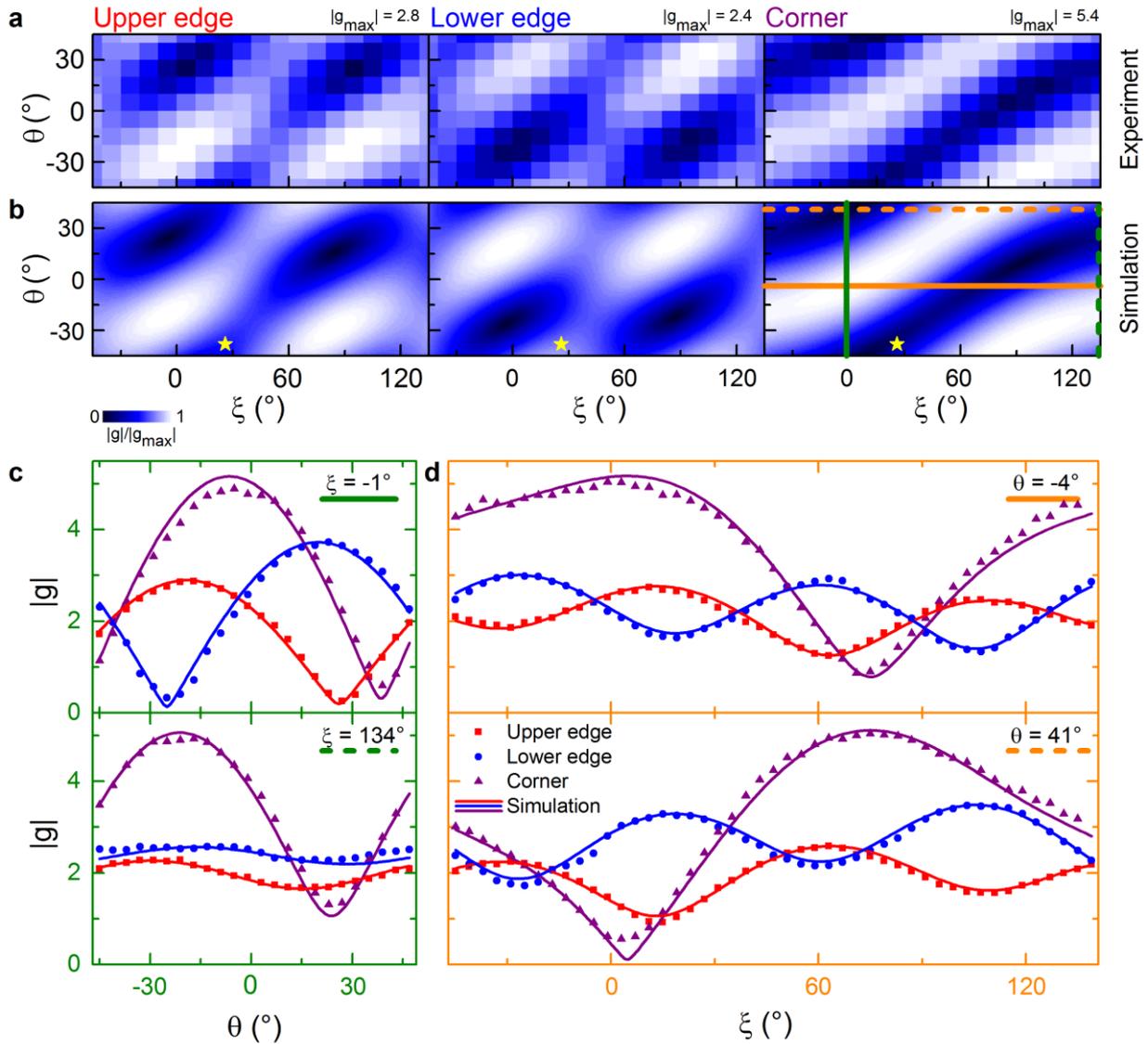

Figure 3: Demonstration of coherent dual interaction for arbitrary polarization states. **a** Coupling constant from experimental electron energy spectra measured for individual (left, middle) and combined (right) near-field actions. **b** Corresponding simulations employing experimentally determined near-field responses $\boldsymbol{\alpha}_{1,2}$ (left, middle) and their coherent sum (right). Yellow star: settings for raster scan in Fig. 1**d**. **c**, **d** Higher resolved lineouts of experimental coupling constant $|g|$ (symbols) and model prediction (solid lines). Position of lineouts indicated by dashed and solid lines in **b**.

To demonstrate a modulation of the total coupling constant $g_{tot}$ by mere manipulation of the relative phase of the interactions $g_1$ and $g_2$, we vary the incident polarization state in such a way as to keep the projections onto the vertical $(\sim|g_1|)$ and horizontal $(\sim|g_2|)$ axes fixed. This is the case for all elliptical polarization states with main axes rotated by 45° with respect to the edges, including ±45° linear as well as left- and right-hand circular polarizations. Figures 2**c** and **d** display the nearly constant coupling strengths at the individual edges resulting from this pure phase variation. Placing the beam at the corner, however, such that it sequentially interacts with both near-fields, we find a strong change in the spectral width of the final electronic state upon a variation through the same set of polarization states (Fig. 2**e**). This conclusively demonstrates the quantum coherence and thus reversibility of the two subsequent interactions. Specifically, a strong recompression of the spectrum is achieved near $\theta$ = 39°. In the spatial wavefunction picture, this corresponds to a cancellation of the initially imprinted phase modulation by the second interaction. The effect of sequential coherent interactions can also be illustrated by spatial maps, in which the total coupling constant is displayed as a function of beam position near the nanostructure (Fig. 1**d**). The individual edges exhibit largely homogeneous coupling constants decaying over a distance of about 100 nm away from the edge (orange regions). In addition, for a destructive relative phase of the individual interactions, a substantially reduced total coupling constant is evident near the corner, at which the electrons traverse both near-fields (green dashed circle).

In order to identify the individual near-field responses, we map the interaction strength for arbitrary incident polarization states by a systematic variation of both wave plate angles. Figure 3**a** displays the measured coupling constants $g_1$, $g_2$, and $g_{tot}$, with higher-resolved lineouts in Figs. 3**c,d** (symbols). For the individual edges (left and middle in Fig. 3**a**, red and blue symbols in Figs. 3**c,d**), we obtain quasi-polarizabilities $\boldsymbol{\alpha_{1,2}} = \alpha_{1,2}\boldsymbol{n_{1,2}}$ with the normalized projection vectors $\boldsymbol{n_1} \cong \begin{pmatrix} 0.09+0.14i \\ 0.99 \end{pmatrix}$ and $\boldsymbol{n_2} \cong \begin{pmatrix} 0.99 \\ -0.05-0.07i \end{pmatrix}$,[37] close to the design aim of $\begin{pmatrix} 0 \\ 1 \end{pmatrix}$ and $\begin{pmatrix} 1 \\ 0 \end{pmatrix}$, and amplitude prefactors of $\alpha_1$=35 (V/nm)$^{-1}$ and $\alpha_2$=44 (V/nm)$^{-1}$. While the vectors $\boldsymbol{n_{1,2}}$ are universal and spatially independent for each of the edges, the specific prefactor sensitively depends on the particular distance from the respective surface. Employing amplitudes $\alpha_1$=52 (V/nm)$^{-1}$, $\alpha_2$=29 (V/nm)$^{-1}$ and a constant phase offset $\varphi_0$ = 1.30, the entire set of measurements near the corner of the structure is successfully described by a summation $g_{tot} = (\boldsymbol{\alpha_1} + e^{i\varphi_0} \cdot \boldsymbol{\alpha_2}) \cdot \boldsymbol{J}$, again clearly demonstrating the phase-controlled quantum coherent interaction with both near-fields. Minor deviations, for example in the incomplete spectral recompression near the minima of $g_{tot}$, are attributed to a spatial average over near-field strengths across the electron beam (see Supplementary Material S4). This leads to small residual sideband populations and highlights the importance of carrying out such experiments with low emittance electron beams, as performed here, using nanotip sources. Dispersive reshaping of the wavefunction, on the other hand, can be excluded for the given spatial separation of the interaction planes (see Supplementary Material S2).

A comment should be made about the invoked phase offset $\varphi_0$. The precise polarization state, at which maximum recompression occurs, is governed by the phase relation between the optical far-field and the respective near-fields, and the phase lag arising from the electron and light propagation between the two interaction planes. Although these phases are physically distinct, in practice, they can be combined in the single phase offset $\varphi_0$, which is sufficient to account for all experiments. For the present measurements, we identify this phase with a precision that corresponds to a timing uncertainty of few attoseconds. This implies a sensitivity of the scheme to phase or timing changes to the free electron wavefunction of this very same magnitude, rendering the presented interferometer an ideal tool to study excitation-induced phase shifts in new forms of electron holography employing the longitudinal degree of freedom. Utilizing this approach to imprint phase information onto the electron wavefunction could be translated to attosecond temporal resolution by, e.g., energy-resolved electron diffraction.

In conclusion, we demonstrated the coherent manipulation of free electron superposition states in sequential near-field interactions. Nanostructures with polarization anisotropy have proven very useful to exert phase-locked dual interactions, analogous to the Ramsey method. The absence of efficient decoherence mechanisms in vacuum renders free-electron wavepackets an ideal system for coherent control schemes, which can be extended to multi-color approaches and additional interaction stages. Other future experiments may utilize this type of "electron-light interferometer" by inserting optically excited materials in the gap for precision measurements of electronic dephasing with sub-cycle resolution. Various further applications include phase-resolved near-field imaging, possible quantum computation schemes using free electrons, or the tailored structuring of electron densities in accelerator beamlines with attosecond accuracy.

**Methods**

The experiments were performed in a recently developed ultrafast transmission electron microscope, featuring a nanoscale photoemitter as a pulsed electron source for electron pulses with high spatial coherence. Specifically, ultrashort electron pulses are generated by localized photoemission from a ZrO/W tip emitter, accelerated to a kinetic energy of 120 keV and focused tightly in close vicinity to a nanostructure. Electron spot diameters down to 3 nm and pulse durations as short as 300 fs were achieved. A SEM image of the nanostructure design is shown in Fig. 1**b**. The two plates with a distance of 5 μm were FIB milled from a single, annealed gold wire (30 μm diameter). The experimental scenario is sketched in Fig. 1**c**: A pump laser beam (800 nm wavelength, dispersively stretched to a pulse duration of 3.4 ps, 250 kHz repetition rate, 23 mW average power) passes a half and a quarter wave plate for polarization control and is focused onto the sample to a spot diameter of about 50 μm (full-width-at-half-maximum). The electron kinetic energy spectra are recorded with an electron energy loss spectrometer (EELS).

## Acknowledgments

We gratefully acknowledge funding by the Deutsche Forschungsgemeinschaft (DFG-SPP-1841 "Quantum Dynamics in Tailored Intense fields", and DFG-SFB-1073 "Atomic Scale Control of Energy Conversion", project A05). We thank S. V. Yalunin for useful discussions, and M. Sivis for help in sample preparation.

## Author contributions

K.E.E. prepared the nanostructure, conducted the experiments with contributions from A.F., and analyzed the data. The manuscript was written by K.E.E. and C.R., with contributions from S.S.. C.R. and S.S. conceived and directed the study. All authors discussed the results and the interpretation.

## Supplementary Material

## S1 Sinusoidal phase modulation

To obtain the electron wavefunction ψ(z,t) after interaction with the optical near-fields, we apply the scattering (S-matrix) approach in the interaction picture (see also Ref. 8). The final wavefunction is given by $|\psi(z,\infty)\rangle = S|\psi(z,-\infty)\rangle$ with the time ordered unitary operator

$$S = T \exp\left(-\frac{1}{\hbar}\int_{-\infty}^{\infty} H_{int} dt\right), \tag{S1}$$

and the interaction Hamiltonian

$$H_{int} = -veA(z,t), \tag{S2}$$

where $v$ is the electron velocity and $e$ the electron charge. The vector potential A(z,t) for the two near-fields separated by the distance L is given by

$$A(z,t) = \frac{F_1(z)}{\omega}\sin(\omega t) + \frac{F_2(z-L)}{\omega}\sin\left(\omega\left(t - \frac{L}{v}\right) - \varphi\right). \tag{S3}$$

φ denotes the phase lag of the second near field induced by the optical path length difference of the driving laser field (corresponding to $d_l$ in Fig. S3). For the wavefunction after interaction we obtain

$$\begin{aligned}
\psi(z,t) &= \exp\left(\frac{iev}{\hbar}\int_{-\infty}^{t} A(z+v\tau,\tau)d\tau\right)\psi(z,-\infty) \\
\Rightarrow \psi(z,+\infty) &= \exp\left(2i|g_1|\sin\left(\frac{\omega z}{v} + \arg(g_1)\right)\right. \\
&\quad \left. + 2i|g_2|\sin\left(\frac{\omega z}{v} + \arg(g_2) + \varphi\right)\right)\psi(z,-\infty). \\
\Leftrightarrow \psi(z,+\infty) &= \exp\left(2i\,\mathrm{Im}\left(e^{i\frac{\omega}{v}z}(g_1 + e^{i\varphi}g_2)\right)\right)\psi(z,-\infty).
\end{aligned} \tag{S4}$$

In the second step, we introduced the coupling constant $g = \frac{e}{2\hbar\omega}\int_{-\infty}^{\infty} F(z)\exp(-i\Delta k z)dz$, as in Ref. 26. It is proportional to the spatial Fourier component of the near-field F(z) along the electron trajectory at the spatial frequency $\Delta k = \omega/v$, which corresponds to the momentum change of an electron at velocity $v$ gaining or losing an energy $\hbar\omega$. Equation S4 evidences that the interaction of the free electrons with the two optical near-fields is describable as a single sinusoidal phase modulation of the electron wavefunction and that the two consecutive interactions coherently add up in the way stated in eq. 3 in the main text.

## S2 Influence of dispersion

Between the two interaction regions, the electron wave function propagates in free space. The momentum-dependent propagation operator is given by

$$T(p) = \exp\left(-\frac{i}{\hbar}\left(\frac{m_e c^2}{\gamma} + \frac{\Delta p^2}{2\gamma^3 m_e}\right)t\right), \tag{S5}$$

where γ is the Lorentz factor and *Δp* the momentum change due to the interaction with the optical near-field, with *Δp* given by *Nℏω/v* for the $N^{th}$-order sideband. During free propagation, the sideband orders acquire different phases, which leads to a dispersive reshaping of the electron wavefunction and, at a certain propagation distance, to a temporal focusing into a train of attosecond pulses.[8] In the present study, the propagation distance is much shorter than the distance to the temporal focus (typically mm-scale). Specifically, the experimental parameters employed here (coupling constants g ≈ 5, propagation distance L = 6 µm, v = 0.6 c, γ = 1.24) yield very small sideband dependent phase shifts on the order of $N^2$·0.17 mrad, such that dispersive effects are negligible.

### S3 Coordinate system and geometric phase offset

The angles θ and ξ define the orientation of the fast axes of the wave plates relative to the horizontal x-axis of the coordinate system, which is indicated by black arrows in Fig. S3**a**. The wave plate setting θ = ξ = -45°, e.g., yields linear laser polarization at -45° to the x-axis.

In the following, we discuss the influence of the sample and beam geometry on the constant phase offset $\varphi_0$. The difference in electron group and laser phase velocity leads to a phase lag, which can be calculated as follows: For a given plate distance *d*, the electron and laser path lengths are $d_e = d/\cos\alpha$ and $d_l = d\cos\beta/\cos\alpha$, respectively, where α = 37° is the sample tilt angle and β = 55° the angle between laser and electron beam. The path length difference corresponds to a timing difference of

$$\Delta t = \frac{d_e}{v_e} - \frac{d_l}{c} = \frac{d}{\cos\alpha \cdot c}\left(\frac{1}{0.6} - \cos\beta\right), \tag{S6}$$

with *v* = 0.6 *c*. A small variation of α by about 2.4° shifts *Δt* by a quarter laser period, i.e. $\varphi_0$ by 90°. Sample tilting thus presents a convenient way to externally control $\varphi_0$. Note that the data displayed in Fig. 2**e** and 3**c** in the main text were recorded at two slightly different sample tilts, resulting in a relative phase shift of $\Delta\varphi_0$ = 81.5°.

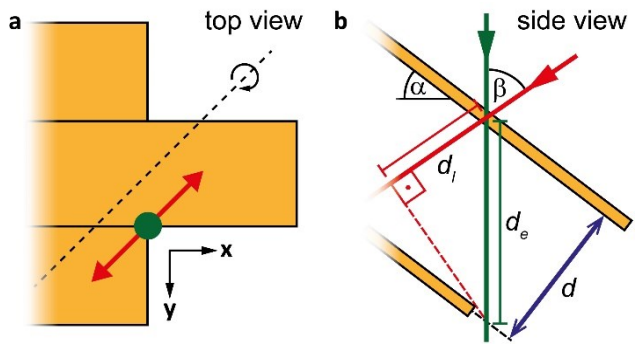

Figure S3: Sample and beam geometry. **a** Top view. Black arrows: coordinate system. Red arrow: incident polarization for wave plate angles θ = ξ = -45°. **b** Side view. The laser beam (red) is focused onto the sample at an angle of β = 55° with respect to the electron beam (green). The sample is tilted by α = 37° (around the dashed black line in **a**). The plate distance *d* = 5 µm determines electron and laser path lengths, $d_e$ and $d_l$, respectively.

## S4 Determination of coupling constant and spatial averaging

In principle, the coupling constants can be inferred from the cutoff energy of the electron energy spectra, which is given by $2|g|\hbar\omega$.[8] For a more precise determination, we extracted coupling constants from a fit of Bessel amplitudes to the data, according to eq. 1. Due to the finite electron beam size, a small spatial average over different coupling constants needs to be taken into account, for which we adopt a Gaussian distribution of the electron intensity in the beam. At the gold edges, the near-field strength can be regarded as homogeneous in directions parallel to edge, and exponentially decaying along the perpendicular direction (cf. Fig. 4**g**). In this case, the probability distribution of coupling constants is given as

$$P(g) \propto \frac{1}{g} \cdot \exp\left(-\frac{1}{2}\left(\frac{l}{\Sigma}\ln\left(\frac{g_0}{g}\right)\right)^2\right), \quad (S7)$$

where $g_0$ is the expectation value of the coupling constant, $l$ the decay length of the near-field strength and $\Sigma$ the electron beam width (standard deviation). For the analysis, we consider a constant ratio $l/\Sigma$ for each near-field.

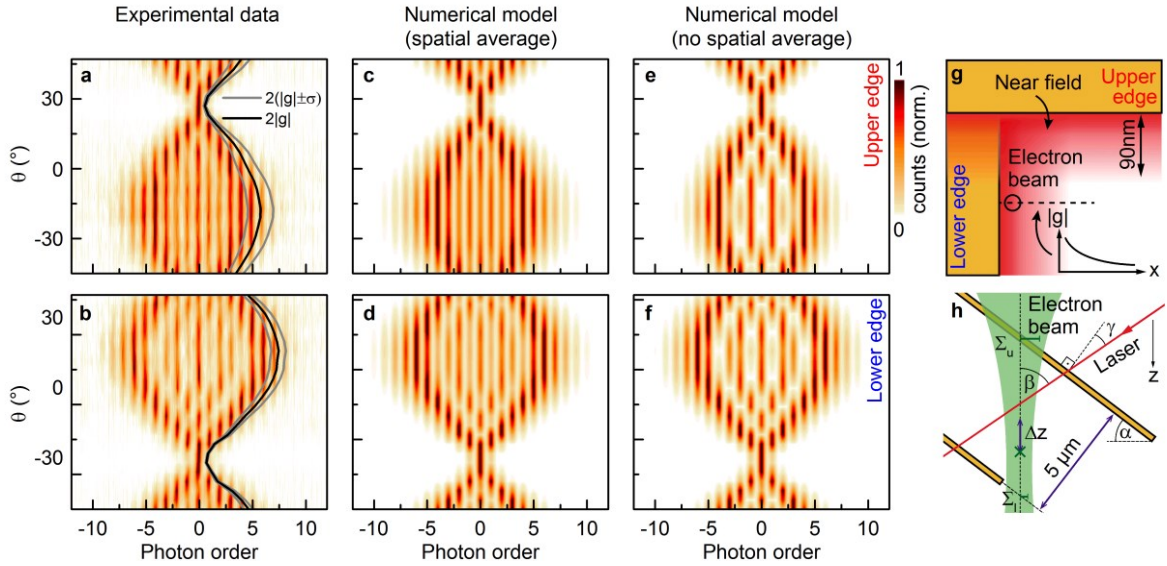

Figure S4: Determination of coupling constant and spatial averaging. **a,b** Experimental electron spectra (normalized to the maximum count rate for clarity) recorded at the upper and lower edge for varying half wave plate angles θ (quarter wave plate at ξ=-1°). **c-f** Bessel amplitudes adapted to the experimental data (with and without spatially averaged coupling constants, respectively). Black and gray curves in **a,b**: Expectation value of coupling constant and its standard deviation (linearly depending on $|g|$, $\sigma_U$ = 0.21|g| and $\sigma_L$ = 0.09|g|). **g** Sketch of the experimental geometry (top view). The coupling constant decays exponentially along the black dashed line (decay length l ≈ 90 nm). All coupling constants within the electron beam (black circle) contribute to the spectra. **h** Sketch of the experimental situation (side view). For experimental angles $\alpha$ and $\beta$ see Fig. S3.

When averaging is taken into account, the experimental data are well reproduced. A comparison of Figs. S4**c,e** illustrates that spatial averaging only weakly affects the visibility of quantum coherent features in the electron energy spectra (cf. Ref. 8). The spectra recorded at the upper edge show stronger averaging compared to the lower edge, since the electron focus is not perfectly centered between the two edges (small displacement $\Delta z$). For the dataset shown here, we obtain $l/\Sigma_U$ ≈ 5 and $l/\Sigma_L$ ≈ 10. Together with the near-field decay length of $l$ ≈ 90 nm (determined from the raster scan in Fig. 1**d**), we find $\Sigma_U$ = 18 nm and $\Sigma_L$ = 9 nm, in accordance with the electron focal spot diameter of 8 nm used in the experiment.